\documentclass{moriond}
\usepackage{graphicx, subfigure}
\usepackage{color,amsmath,amssymb}
\usepackage{txfonts}
\usepackage[usenames,dvipsnames]{xcolor}
\usepackage{hyperref}
\usepackage{wrapfig}

\newcommand{\Mpc}{\,\mathrm{Mpc}}

\def\be{\begin{equation}}
\def\ee{\end{equation}}
\def\bea{\begin{eqnarray}}
\def\eea{\end{eqnarray}}

\newcommand{\Photo}{}

\begin{document}
\vspace*{4cm}
\title{LENSING DISPERSION OF SNIa AND SMALL SCALES OF THE PRIMORDIAL POWER SPECTRUM}

\author{IDO BEN-DAYAN}

\address{Deutsches Elektronen-Synchrotron DESY, Theory Group, D-22603 Hamburg, Germany}

\maketitle\abstracts{ 
Probing the primordial power spectrum at small scales is crucial for discerning inflationary models, especially if BICEP2 results are confirmed. We demonstrate this necessity by briefly reviewing single small field models that give a detectable gravitational waves signal, thus being degenerate with large field models on CMB scales. A distinct prediction of these small field models is an enhancement of the power spectrum at small scales, lifting up the degeneracy. We propose a way to detect this enhancement, and more generally, different features in the power spectrum at small scales $1\lesssim k \lesssim 10^2-10^3 \Mpc^{-1}$ by considering the existing data of lensing dispersion in Type Ia supernovae. We show that for various deviations from the simplest $n_s\simeq 0.96$ the lensing dispersion cuts considerably into the allowed parameter space by PLANCK and constrains the spectrum to smaller scales beyond the reach of other current data sets.}

\section{Introduction}
State of the art CMB and $\mathrm{Ly_\alpha}$ measurements probe only about $8$ e-folds, $(H_0\lesssim k \lesssim 1 \Mpc^{-1})$ out of the expected $60$ e-folds of observable inflation \cite{Ade:2013uln}, rendering a huge degeneracy between inflationary models. Even a confirmation of the BICEP2 measurement \cite{Ade:2014xna}, will not break all the degeneracy. For example, small field models with a non-monotonic $\epsilon$  reproduce a spectrum similar to that of a monomial $V\sim \phi^n$ for a limited range of wave numbers \cite{BenDayan:2009kv}. Even within the class of large field models there is a degeneracy that can only be lifted by probing enough e-folds of the power spectrum. 
The answer lies in probing smaller scales of the power spectrum. In \cite{Ben-Dayan:2013eza} we proposed using the lensing dispersion of type Ia supernovae as a novel cosmological probe and specifically as a constraint on the primordial power spectrum at small scales. See also \cite{Hamana:1999rk}. %, Minty:2001jg, Dodelson:2005zt, Amendola:2013twa, Quartin:2013moa, Fedeli:2013yfa, Castro:2014oja}.
The lensing dispersion, $\sigma_{\mu}$ is sensitive to  $0.01 \lesssim k \lesssim 10^2-10^3 \Mpc^{-1}$, thus giving access to $2-3$ more decades ($4-7$ e-folds) of the spectrum, even using only \textit{current data}.

In the next section, we review the non-monotonic $\epsilon$ idea. In section $3$ we present how the lensing dispersion probes the primordial power spectrum on small scales. Section $4$ describes the results for various parameterizations of the spectra, complementing \cite{Ben-Dayan:2013eza}, and some discussion.

\section{Small field models and large r}
Consider canonically normalized, single field models $V(\phi)=\Lambda^4\sum_{n=0}a_n\phi^n$, assuming CMB scales are at $\phi\simeq 0$. Generically $a_0$ sets the scale of inflation, $a_1$ sets the tensor to scalar ratio $r=16\epsilon$, $a_2$ sets $n_s$ etc. A small field model $\Delta \phi <1$
requires parametric tuning of a few parameters for a successful model of inflation, i.e. $\epsilon,|\eta| \ll1$, to get $60$ e-folds and $n_s\simeq 0.96$. \footnote{The virtue of small field models, i.e. parametric tuning of only a few operators is especially relevant if one considers an inflation as a fundamental field. For effective low energy degrees of freedom, one can get $\Delta \phi \gg1 $ in a rather natural way \cite{Kim:2004rp}.}. %,Ben-Dayan:2014zsa,Tye:2014tja,Ben-Dayan:2014lca} [DI?].}. 
 This generically means $a_1\ll1$ and hence $r\ll 0.01$ in odds with the BICEP2 result \footnote{In \cite{BenDayan:2009se}, I erroneously claimed a non-monotonic $\epsilon$ evades the Lyth bound from 1996 \cite{Lyth:1996im}, contributing to confusion in the literature. This was promptly corrected in \cite{BenDayan:2009kv}. The strict '96 bound cannot be evaded, only the BL bound \cite{Boubekeur:2005zm}, which assumes a monotonic $\epsilon$.}. A large field generically means functional tuning, for example $a_{n\neq2}=0$, for all $n$, which gives a free massive inflaton. Such models give $r\sim 0.1$, in accord with the BICEP2 findings. One would like a UV theory that explains the functional tuning we use. Moreover, taken at face value, the $r=0.2$ BICEP2 result is in tension with PLANCK, unless the cosmological model is further extended to include primordial Helium, additional light degrees of freedom or a scale dependent spectral index $n_s(k)$, in a way that suppresses the power on intermediate scales. Regardless, hints of "running" $\alpha(k_0)\equiv dn_s/d \ln k$ have been around since WMAP1 \cite{Peiris:2003ff}. 

\begin{figure}
%\begin{minipage}{0.33\linewidth}
\centerline{\includegraphics[height=4cm,width=0.7\linewidth]{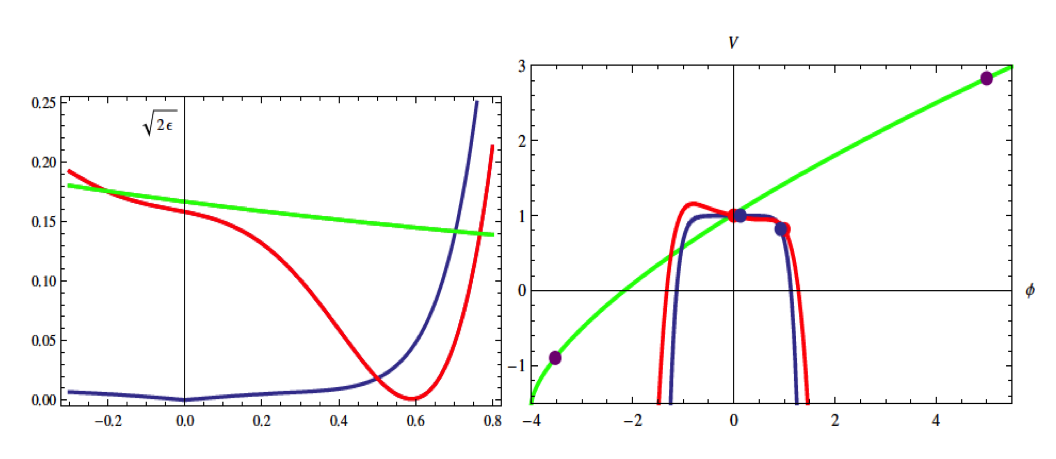}}
\caption{$\sqrt{2\epsilon(\phi)}$ (left panel) and $V(\phi)$ (right panel) for typical hilltop (blue), monomial $V\sim \phi^{2/3}$ with a constant shift for clarity (green) , and the non-monotonic $\epsilon$ models (red).}%, reproduced from \cite{BenDayan:2009kv}.}\label{nonmonotonic}
%\end{minipage}
%\vspace{-1cm}
\end{figure}
Several years prior to PLANCK and BICEP2, in \cite{BenDayan:2009kv} we demonstrated the key idea, that non-monotonic $\epsilon=\frac{1}{2}\left(\frac{V'}{V}\right)^2$ allows small field models to have $r\sim 0.1$, avoiding the need of functional tuning, which is especially interesting in light of BICEP2 \cite{Choudhury:2014kma}.
%The main idea is a non-monotonic $\epsilon=\frac{1}{2}\left(\frac{V'}{V}\right)^2$. 
 If at CMB scales, $\epsilon$ is rather large, then from $r=16\epsilon$ we get detectable signal, $r\sim 0.1$. However, away from the CMB scales, $\epsilon$ decreases, giving many e-folds of inflation, $N=\int d\phi/\sqrt{2\epsilon}$. In Figure 1, reproduced from \cite{BenDayan:2009kv}, we demonstrate the behaviour of $\sqrt{2\epsilon}=|V'/V|$ and the potential $V$ as a function of the inflation.
One can have an arbitrary number of e-folds in a very small interval $\Delta \phi \ll 1$  \cite{BenDayan:2009kv,Itzhaki:2008hs}. %, especially if the end of inflation comes from a hybrid mechanism [Hebecker]. 
%Therefore, the main limitation is not the number of e-folds, but 
Therefore, the main limitation is the scale dependence of the power spectrum, $P\sim V/\epsilon$, since by now about $8$ e-folds have been measured with limited amount of scale dependence parameterized by $\alpha(k_0)\equiv dn_s/d \ln k, \beta(k_0)\equiv d^2 n_s/d \ln k^2$.
  
Because of the non-monotonic $\epsilon$, a distinct \textit{prediction} of the models, which was made prior to PLANCK and BICEP2 results, is the enhancement of the power spectrum at small scales. %Specifically, the notion of enhanced scale dependence as PLANCK and BICEP2 may suggest is a prediction, and not an additional fitting of the model. 
 In \cite{BenDayan:2009kv} the spectrum was calculated numerically by solving the Mukhanov-Sasaki equation, and in \cite{Chluba:2012we} it was argued that a spectrum with a bend at some $k_i$ is a good approximation of the model, which we will use in section $4$. Knowing the power spectrum at smaller scales is interesting by itself for a better understanding of inflation. Specifically, it can break the degeneracy between the above models and the monomial ones, because the former will have enhanced power at small scales \footnote{Measuring the spectrum for enough e-folds we will be able to discriminate even between the simplest models via spectral distortions \cite{Chluba:2013pya}, and perhaps even get a hint of a stringy origin \cite{Ben-Dayan:2013fva}.}. We therefore suggest the lensing dispersion of SNe as a probe of the small scale power spectrum and a novel cosmological probe in general.% and one should seek a better measurement of the small scale power spectrum 

\section{Lensing Dispersion of SNIa}

%\begin{figure}[t]
%\includegraphics[width=7cm]{fig1.pdf}
%\caption{Log-Log plot of the ``transfer functions'' (\ref{transfer_function}) at redshift $z=1$ multiplied by $c=1$ (solid black), $c=0.5$ (dashed grey), $c=0.1$ (dashed red) $c=0.01$ (dashed green). Solid blue, cyan and purple curves are the step functions in (\ref{step_funct}) with $b=3, k_{NL}=1\Mpc^{-1}$; $b=10, k_{NL}=2 \Mpc^{-1}$; $b=50, k_{NL}=15 \Mpc^{-1}$, respectively.}\label{TLNL}
%%\vspace{-0.5cm}
%\end{figure}

Using the light-cone averaging approach up to second order in the Poisson (longitudinal) gauge \cite{Gasperini:2011us}, %BenDayan:2012pp, BenDayan:2012ct, BenDayan:2012wi,  BenDayan:2013gc}, 
 a simple expression for the lensing dispersion in \cite{Ben-Dayan:2013eza} was derived: 
\bea
\label{sigmuLNL}
\sigma_{\mu}^2 &\simeq& \left(\frac{5}{ \ln 10}\right)^2 \frac{\pi}{\Delta\eta^2} \int_{\eta_s^{(0)}}^{\eta_o} \frac{d\eta_1 dk}{k}P_{\Psi}(k,\eta_1)k^3(\eta_1-\eta_s^{(0)})^2(\eta_o-\eta_1)^2,\,\\
 \label{dimless} &\simeq&  \left(\frac{5}{ \ln 10}\right)^2 \frac{\pi}{\Delta \tilde \eta^2} \left(\frac{k_{eq}}{H_0}\right)^3 \int d\tilde \eta_1 dp P_{\Psi}(p,\tilde \eta_1)p^2 (\tilde \eta_1-\tilde \eta_s^{(0)})^2(\tilde \eta_o-\tilde \eta_1)^2.
 %& ~~~~~~~~~~~~~~~~~~~~~~~~~~~~\,,\nonumber
\eea
% \left(\frac{5}{ \ln 10}\right)^2
%where the gravitational potential $\Psi=\Psi(\eta_i,r_i,\tilde\theta^a)$ is evaluated along the past light-cone at $r_i=\eta_o-\eta_i$, 
 where $\eta_o$ is the observer conformal time, $\eta_s^{(0)}$ is the conformal time of the source with unperturbed geometry, $\Delta \eta(z)=\eta_o(z)-\eta_s^{(0)}(z)=\int^z_0
\frac{dy}{H_0\sqrt{\Omega_{m0}(1+y)^3+\Omega_{\Lambda0}}}$,
and $P_{\Psi}$ is the linear (LPS, $P_L$) or non-linear dimensionless power spectrum (NLPS, $P_{NL}$) of the \textit{gravitational potential}. In the second line we switched to dimensionless variables, $\tilde \eta= H_0 \eta$ and $p=k/k_{eq}$ \footnote{The choice of the equality scale $p=k/k_{eq}$ is because we know the general behaviour of $P_L$, or more precisely, its transfer function $T(k)$ which is constant for $p<1$ and scales like $p^{-2} \ln p$ for $p \gg 1$}. 
Equation \eqref{dimless} demonstrates the relevant physical scales $H_0$ and $k_{eq}$, the sensitivity to scales smaller than the equality scale $p>1$, and the expected enhancement pattern $(k_{eq}/H_0)^3$ in the linear regime and potentially additional $(k_{NL}/k_{eq})^3$ at a redshift dependent non-linearity scale, $k_{NL}$.
So we have a direct probe of the integrated late-time power spectrum and of the cosmological parameters. 
 
At redshift $z\sim 1$ the dispersion, $\sigma_{\mu}$, grows approximately linearly with redshift, so the best constraints will be obtained from the maximal available redshift of current data, $z=1$. We do not have a definite detection, but a conservative $2$-sigma upper bound $\sigma_{\mu}(z=1)\leq 0.12$ \cite{Jonsson:2010wx}. It is conservative because all analyses \cite{Gasperini:2011us,Jonsson:2010wx,Holz:2004xx,Betoule:2014frx} point to a lower value of the dispersion, at most $\sigma_{\mu}(z)\simeq0.093z$, \cite{Holz:2004xx}. %Moreover, the recent JLA analysis [JLA], suggests that even the \textit{total dispersion} at $z\simeq 1$ is only $\sigma_{\mu} \simeq 0.12$.  
Moreover, the most up-to-date JLA analysis uses the actual value from \cite{Jonsson:2010wx}$, \sigma_{\mu}=0.055 z$ and still sees a decrease as a function of redshift in the left over 'coherent' or 'intrinsic' dispersion, suggesting that even the \textit{total dispersion} at $z\simeq 1$ is only $\sigma^{tot}_{\mu} \lesssim 0.12$, \cite{Betoule:2014frx}. Additionally, partial sky coverage and higher redshift SN, which have already been used for cosmological parameter inference, will increase the dispersion, making our analysis even more conservative.
%and other analyses which imply a much smaller dispersion \cite{Holz:2004xx, Kronborg:2010uj, BenDayan:2013gc, Karpenka, March, Conley11, JLA}, none unfortunately proclaiming a definite detection.
% In general, the $k^2$ enhancement makes $\sigma_{\mu}$ a sensitive probe to the small scales of the power spectrum. To make this claim transparent, let us switch to dimensionless variables, $\tilde \eta= H_0 \eta$ and $p=k/k_{eq}$ \footnote{The choice of the equality scale $p=k/k_{eq}$ is because we know the general behaviour of $P_L$, or more precisely, its transfer function $T(k)$ which is constant for $p<1$ and scales like $p^{-2} \ln p$ for $p \gg 1$}:
% \begin{eqnarray}
%\label{dimless}
% \sigma_{\mu}^2 \simeq \left(\frac{5}{ \ln 10}\right)^2 \frac{\pi}{\Delta \tilde \eta^2} \left(\frac{k_{eq}}{H_0}\right)^3 \int d\tilde \eta_1 dp P_{\Psi}(p,\tilde \eta_1)p^2 (\tilde \eta_1-\tilde \eta_s^{(0)})^2(\tilde \eta_o-\tilde \eta_1)^2\,.
% \end{eqnarray}
\begin{figure}[t]
%\begin{minipage}
\hspace{-0.5cm}
\subfigure{\includegraphics[width=7.3cm,height=73mm]{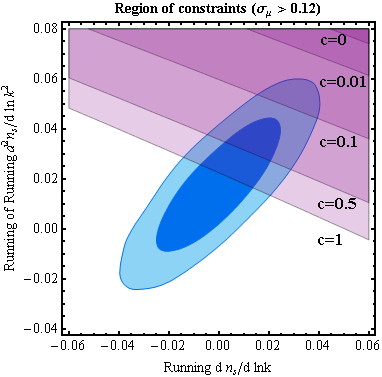}}\quad \hspace{0.2cm}
\subfigure{\includegraphics[width=7cm]{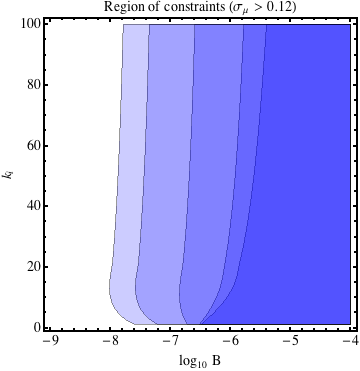}}
\vspace{-0.3cm}
\\
%\caption{Regions of allowed parameters combined with  PLANCK data. The ellipses are the 68\% and 95\% CL contours from \cite{Ade:2013uln}. In the colored regions $\sigma_{\mu}(z=1)>0.12$ and are disfavoured for $b=0,1,3,10,50$ (left panel) and $c=0,0.01,0.1,0.5,1$ (right panel)}\label{planck_constr}
%\end{figure*}
%\begin{figure*}
\subfigure{\includegraphics[width=7cm]{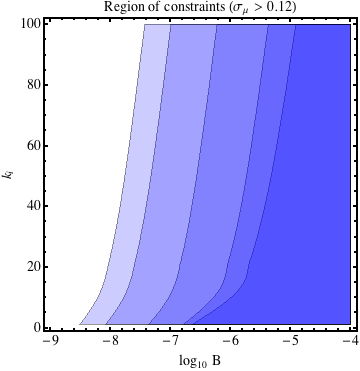}}\hspace{0.5cm}
\subfigure{\includegraphics[width=7cm]{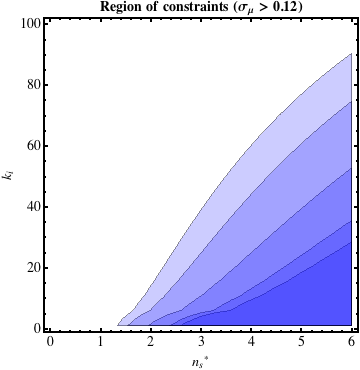}}
%\vspace{-0.3cm}
%\end{minipage}
\caption{Exclusion plots for the different parameterizations. Shaded regions correspond to $\sigma_{\mu}(z=1)>0.12$ with $c=0,0.01,0.1,0.5,1$ respectively from dark to light and are disfavoured. The different parameterizations are: top left eq. \eqref{eq:run}, top right eq. \eqref{eq:bump}, bottom left  eq. \eqref{eq:step}, bottom right  eq. \eqref{eq:bend}. In the top left panel the ellipses correspond to $68\%$ and $95\%$ likelihood contours from PLANCK.}
\label{planck_constr}
\end{figure}

 The main limitation of \eqref{dimless} is the validity of the spectrum \cite{Gasperini:2011us}, because for $k\gg H_0$ standard cosmological perturbation theory breaks down, and one has to resort to numerical simulations to get an approximate fitting formula for the power spectrum. We use the HaloFit model \cite{Inoue:2012px} with $k_{UV}=320h \Mpc^{-1}$. For the standard case $P_{k}=A_s(k/k_0)^{n_s(k_0)-1}$, $\sigma_{\mu}(z=1, k_{UV}=320h \Mpc^{-1})\simeq0.08$. Within a certain range, varying $H_0,\Omega_{m0},k_{UV}$ can account at most for $15\%$ difference \cite{Ben-Dayan:2013eza}. Hence the bound cannot be saturated by varying the background parameters and/or integrating up to arbitrarily small scales. Hence, it can be used for probing small scales of the power spectrum. %We have verified that varying $k_{UV}\in(30h,\infty) \Mpc^{-1}$ or $H_0$ and $\Omega_{m0}$ independently within the range $H_0\in[64, 70]$, $\Omega_{m0}\in[0.27, 0.36]$ gives at most $15\%$ change in the value of $\sigma_{\mu}$. Taken that , the bound cannot be saturated by varying the background parameters and/or integrating up to arbitrarily small scales. Hence the bound can be useful for scale dependence in the power spectrum. % constraining $\alpha(k_0)$ and $\beta(k_0)$ and 
After fixing all the background parameters, including $A_s,n_s(k_0=0.05 \Mpc^{-1})$ to the most likelihood value of \cite{Ade:2013uln}, we achieve accuracy of about $20\%$.
 
\section{Results}
%In principle, the primordial power spectrum is not limited to a specific parametrization. In practice,
%it is typically parameterized as $P_{k}=A_s(k/k_0)^{n_s(k_0)-1}$, where $k_0$ is a suitable ``pivot scale''. 
We analyze four different, more general parameterizations of the spectrum and the corresponding panel in Figure $2$:
\begin{align}
P_k &=A_s\left(\frac{k}{k_0}\right)^{n_s(k_0)-1+\frac{\alpha(k_0)}{2} \ln\frac{k}{k_0}+\frac{\beta(k_0)}{6} \ln^2\frac{k}{k_0}}\,,\label{eq:run}&\quad& \text{top left panel}\\
P_k&=A_s \left(\frac{k}{k_0}\right)^{n_s(k_0) - 1} + B \left(\frac{\pi e}{3}\right)^{3/2} \left(\frac{k}{k_i}\right)^3 e^{-\pi/2 (k/k_i)^2}\,, \label{eq:bump} &\quad& \text{top right panel}\\
P_k&=A_s \left(\frac{k}{k_0}\right)^{n_s(k_0) - 1} \left[1+\frac{B}{A_s}\Theta(k-k_i)\right]\,, \label{eq:step} &\quad& \text{bottom left panel}\\
P_k&=A_s \left(\frac{k}{k_0}\right)^{n_s(k_0) - 1} \left[\Theta (k_i-k)+\left(\frac{k}{k_i}\right)^{n_s^*(k_0)-1}\Theta(k-k_i)\right]. \label{eq:bend} &\quad& \text{bottom right panel}
\end{align}
where $\Theta$ is the Heaviside function, $\alpha(k_0)\equiv dn_s/d \ln k, \beta(k_0)\equiv d^2 n_s/d \ln k^2$ are the ``running'' and "running of running" of the spectral index, %and $\beta$, the ``running of running''. %The best constraints on $\alpha,\beta$ with $k_0=0.05 Mpc^{-1},\,n_s(k_0) \simeq 0.96$ are given by PLANCK \cite{Ade:2013uln} and $\mathrm{Ly_\alpha}$ \cite{Zhao:2012xw}.
%These analyses are only probing the range $H_0\leq k\leq 1 Mpc^{-1}$. 
\eqref{eq:bump} is a typical parameterization of one episode of particle production \cite{Barnaby:2009dd}, \eqref{eq:step} describes a step in the power spectrum, for instance due to several episodes of inflation \cite{Adams:1997de}, and \eqref{eq:bend} describes an enhancement which is not necessarily captured just by running. The models discussed in section $2$ fit the latter parameterization.  

For the above parameterizations, the HaloFit formula is not reliable anymore due to its sensitivity to initial conditions. %For example, with $\alpha=0.04$ and $\beta=0.05$ the existing HaloFit actually gives enhancement of a few at $1<k<175 \,\Mpc^{-1}$ and a \textit{suppressed} power spectrum compared to the linear one at larger $k$. 
 It is nevertheless obvious that the non-linear evolution causes clustering and enhances the power spectrum. 
 %For example, at redshift $z=1$, the ratio between the HaloFit formula, $P_{NL}(k,z)$, with standard initial conditions $(n_s\simeq0.96,\, \alpha=\beta=0)$  and the linear power spectrum $P_{NL}(k,1)/P_L(k,1)$ is the solid, thick, black curve plotted in Fig.~\ref{TLNL}. %Already at $k=1 \Mpc^{-1}$ the non-linear power spectrum is a factor of a few larger than the linear one, and for $k\geq 10 \Mpc^{-1}$, it behaves as a power law with a scaling exponent of nearly $1/2$.
We therefore define a ratio,
$
F(k,z)\equiv \frac{P_{NL}(k,z)}{P_L(k,z)}\,, \label{transfer_function}
$
where $P_{NL}$ is the non-linear power spectrum, $P_L=(3/5)^2 P_k T^2(k)g^2(z)$ is the linear spectrum, $g(z)$ is the growth factor and $T(k)$ is the transfer function with baryons \cite{Gasperini:2011us}, in the standard scenario $n_s\simeq0.96$. %$n_s\simeq0.96,\alpha=\beta=B=0, n_s^*=1$. 
We take the enhancement into account by substituting in \eqref{sigmuLNL}:
%in two simple ways. The first method is by the Heaviside function $\Theta (k)$. Here we are not limited to the HaloFit formula, so we perform the following substitution in equation (\ref{sigmuLNL}):
%and we evaluate $\sigma_{\mu}$ for $b=0,3,10,50$ with corresponding $k_{NL}=1,1,2,15 \Mpc^{-1}$, such that the step function is always underestimating the transfer function $F$, so this is a very conservative estimate.
%The step functions are the solid blue, cyan and purple lines in Fig.~\ref{TLNL}.
%The second method is to use $F$ of the HaloFit model, such that
\be
P_{\Psi}\rightarrow P_L(k,z)(1-c+c F(k,z))\,, \label{Finterp}
\ee
and evaluate $\sigma_{\mu}$ with $c=0,0.01,0.1,0.5,1$. $c=0$ corresponds to computing the dispersion with the linear power spectrum only, while
$c=1$ corresponds to exactly following the HaloFit enhancement pattern. Except $c=1$ all values of $c$ are underestimates %The resulting enhancement at $z=1$, is plotted in Fig.~\ref{TLNL} as green, red and grey dashed lines. 
 \footnote{In \cite{Ben-Dayan:2013eza}, we also considered a step function of the sort $
P_{\Psi} \rightarrow P_L(k,z)(1+b\,\Theta(k-k_{NL}))\label{step_funct}$, %which are the solid steps in Figure 1, 
for $b=0,3,10,50$ with corresponding $k_{NL}=1,1,2,15 \Mpc^{-1}$ always underestimating the ratio $F$. The results are very similar to that of \eqref{Finterp}.}.
The results are presented in Figure \ref{planck_constr}. In all panels, coloured regions give $\sigma_{\mu}(z=1)\geq0.12$ for $c=0,0.01,0.1,0.5,1$ from dark to light and are disfavoured.

From Fig.~\ref{planck_constr}, it is obvious that the lensing dispersion or its absence is an extremely powerful cosmological probe. Even if a scale dependent spectral index induces clustering which is \textit{an order of magnitude smaller} than the standard constant $n_s$ scenario, some of the parameter space allowed by PLANCK is ruled out. Moreover, the analysis probes the spectrum up to $k\sim320 h\Mpc^{-1}$, more than two orders of magnitude beyond PLANCK's lever arm ($\sim5$ e-folds more). %irrespective of whether models are ruled in or out.  
 Calling $c=0.1$ 'realistic' and $c=1$ 'optimistic', the spectrum never goes above $(6,3.7)\times 10^{-7}$ up to $k\leq 320h\Mpc^{-1}$ respectively %, and for the optimistic case of $c=1$ $A_s=\cdots$ up to $k\leq 320h \Mpc^{-1}$, 
 for features up to $k_i\leq 100 \Mpc^{-1}$. The only exception is \eqref{eq:bend} which gives  $P_k(320h \Mpc^{-1})=2.3 \times 10^{-6}$ in the realistic case for $k_i\leq 50\Mpc^{-1}$. This is due to a slow enhancement and on smaller scales so it is quickly erased via Silk damping. 
We are currently analyzing numerical simulations that will test our claims \cite{IT}. 
Combining SN lensing in analyses (present and forthcoming missions), will undoubtedly allow a much better determination of the cosmological parameters. %It can be treated as a prediction of inflationary models.
 
%In the more realistic case where the enhancement is similar to the HaloFit model, such as $c=0.5,1$,  one gets strong bounds on the allowed parameters, that can be expressed as a linear relation:
%\bea
%\beta(k_0) \leq0.036-0.42\,\alpha(k_0), \quad c=0.5\\
%\beta(k_0)\leq0.022-0.44\,\alpha(k_0), \quad c=1.
%\eea
 %The realistic case of $\beta(k_0)\leq 0.022$
%nicely matches PLANCK's $\alpha(k_0)=0^{+0.016}_{-0.013}\,,\beta(k_0)=0.017_{-0.014}^{+0.016}$.
 %Obviously, a definite detection of lensing will enable a more stringent analysis similar to CMB lensing.

%It is very appealing to add the lensing dispersion constraint to the likelihood analysis of the PLANCK data. We believe that numerical simulations with initial conditions as suggested here, $\alpha(k_0),\beta(k_0)\neq0$, which will give a more accurate late time power spectrum, will yield similar results, thus strengthening our argument. These simulations are already on their way.
%  Last, we have suggested using the (absence of) dispersion to constrain the primordial power spectrum. Since the dispersion depends on several cosmological parameters, it can be useful in constraining other fundamental cosmological parameters as well.

\section*{Acknowledgments}
We thank Tigran Kalaydzhyan for collaborating in the early stage of the work.
This work is supported by the German Science Foundation (DFG) within the Collaborative Research Center (CRC) 676 
``Particles, Strings,
 and the Early Universe''.
%\section*{Appendix}
%
% We can insert an appendix here and place equations so that they are
%given numbers such as Eq.~\ref{eq:app}.
%\be
%x = y.
%\label{eq:app}
%\ee

\section*{References}
%\bibliography{moriond}

\end{document}